\documentclass[sigconf,nonacm]{acmart}

\setcitestyle{numbers,sort&compress}
\usepackage{booktabs}
\usepackage{threeparttable}
\usepackage{subcaption}
\usepackage{tikz}
\usepackage{multirow}
\usepackage{float}
\usepackage{enumitem}
\usepackage{stfloats}
\usetikzlibrary{positioning}
\usepackage{xurl}
\usepackage{hyperref}
\begin{document}

\title{What Drives Students' Use of AI Chatbots? Technology Acceptance in Conversational AI}

\author{Griffin Pitts}
\authornote{Corresponding author.}
\affiliation{%
  \institution{North Carolina State University}
  \city{Raleigh}
  \state{North Carolina}
  \country{United States}
}
\email{wgpitts@ncsu.edu}

\author{Sanaz Motamedi}
\affiliation{%
  \institution{Pennsylvania State University}
  \city{University Park}
  \state{Pennsylvania}
  \country{United States}
}
\email{sjm7946@psu.edu}

\renewcommand{\shortauthors}{Pitts \& Motamedi}

\begin{abstract}
Conversational AI tools have been rapidly adopted by students and are becoming part of their learning routines. This widespread use comes amid ongoing concerns about the reliability of AI-generated responses and the appropriate role of AI in education. Drawing on the Technology Acceptance Model, we examine how students' perceptions of conversational AI, including its usefulness, ease of use, trustworthiness, enjoyment, and social influence, relate to their intentions to use these tools for learning. We investigate both the direct and indirect relationships among these perceptions to understand how they are associated with students' behavioral intentions. Using survey data from 229 students and partial least squares structural equation modeling, we found that perceived usefulness was the strongest direct predictor of intention, followed by perceived enjoyment. Perceived ease of use, trust, and subjective norms did not have significant direct associations with intention. However, trust and subjective norms had significant total indirect associations, while perceived enjoyment had both significant direct and indirect associations. The model explained 70.2\% of the variance in behavioral intention. These findings suggest that students' intentions to use conversational AI are associated less directly with effort-related considerations and more with perceived usefulness and affective experience. They also highlight the importance of understanding what students perceive as valuable in conversational AI interactions, particularly when designing tools intended to encourage engagement while supporting learning.
\end{abstract}


\maketitle

\section{Introduction}
Students have long relied on instructors, peers, textbooks, and online resources for assistance with their academic work. Conversational AI chatbots have introduced another way for students to seek support, offering explanations of course material, writing assistance, and help with problems through natural-language interactions \cite{von2023artificial,pitts2025student,yilmaz2023augmented}. Unlike traditional learning resources that provide relatively fixed content, these systems allow students to ask follow-up questions, request clarification, and receive responses tailored to their inquiries. The growing presence of AI in higher education has been reflected in surveys of student use across different countries. For instance, 63.4\% of 6,311 surveyed university students in Germany reported prior usage of AI tools in their studies, with ChatGPT being the primary tool \cite{von2023artificial}. In Hong Kong, 44.9\% of 399 surveyed students reported frequency of use of generative AI technologies between the options of sometimes, often, and always \cite{chan2023students}. Similarly, in the United States, 64.5\% of 262 surveyed undergraduate students reported usage of AI chatbots at least once per week \cite{pitts2025student}. However, adoption has not been uniform. A study of undergraduate engineering students found that 40.2\% reported limited or no use of generative AI during design projects, with students expressing concerns about output reliability and alignment with project needs \cite{dai2025students}. These findings suggest that while conversational AI chatbots have a more common presence in students' learning environments, patterns of adoption remain heterogeneous and guided by multiple considerations. Given the rapid adoption of conversational AI across educational contexts, identifying the factors associated with students' intention to engage with these systems can guide the future design and deployment of pedagogically-aligned conversational AI technologies. As a first step, prior survey and qualitative studies provide insight into how students describe the benefits and concerns that may inform their usage intentions. 

Students engage with conversational AI across a variety of academic activities, often using these tools independently of structured instructional settings. In writing-intensive contexts, AI chatbots are commonly used to assist with idea generation, text organization, paraphrasing and proofreading \cite{vcrvcek2023writing,von2023artificial,pitts2025student}. Students report using conversational AI systems to support independent learning by clarifying concepts, obtaining simplified explanations of course material, exploring unfamiliar topics, and preparing for testing \cite{von2023artificial,pitts2025student}. In technical disciplines such as programming, AI chatbots help with code generation, debugging assistance, and explanatory feedback, with potential benefits for self-efficacy and motivation~\cite{yilmaz2023augmented,vrechtavckova2025finding,pitts2025llmsurvey,hoq2026explainable}. Educational institutions have also explored integrating AI chatbots into more structured learning environments, including virtual teaching assistants, personalized tutoring systems, and academic support services \cite{okonkwo2021chatbots,labadze2023role}. Across these applications, conversational AI differs from many conventional educational technologies through its ability to support natural-language dialogue, generate responses to individual questions, and accommodate follow-up interactions. 

Student perspectives of AI in education vary. Students frequently cite benefits such as access to personalized feedback, support for studying and exam preparation, instructional guidance, and expanded access to knowledge resources \cite{von2023artificial,pitts2025student}. At the same time, concerns remain widespread. Academic integrity is a recurring theme, with students expressing anxiety about plagiarism, unfair advantage, and difficulty verifying authorship \cite{arslan2025ai,pitts2025student}. Others raise concerns about overdependence on AI tools and the potential erosion of critical thinking, problem-solving, and writing skills \cite{pitts2025student,pitts2025reliance,qiao2025use}. Questions of accuracy and reliability are also salient, as AI-generated responses may contain errors, fabricated references, or misleading explanations \cite{pitts2025student,pitts2025reliance,qiao2025use}. Broader ethical considerations, including data privacy risks, algorithmic bias, and environmental impact, further complicate students' evaluations of these technologies \cite{pitts2025student,kasneci2023chatgpt}. Despite these concerns, many students view AI literacy as essential for future careers and believe that learning to use AI tools effectively will be an important professional skill \cite{von2023artificial,chan2023students}. These mixed perceptions raise the question: what drives students to adopt AI chatbots for learning despite ongoing uncertainty and perceived risks associated with the technology?

The Technology Acceptance Model (TAM) has been widely applied to understand technology adoption and emphasizes perceived usefulness and perceived ease of use as determinants of behavioral intentions \cite{davis1989technology}. Prior studies demonstrate the value of TAM for examining students' adoption of digital learning technologies \cite{bailey2024integrating,liu2025intention,chintalapati2017examining,pitts2025lessons,lin2025learner,choung2023trust,wu2017continuance}. Recent research has also applied extended technology acceptance frameworks to examine students' attitudes, intentions, and use of generative AI tools (e.g., \cite{polyportis2025understanding}). Conversational AI systems differ from many previously studied technologies through their natural-language interactions and probabilistic outputs, often functioning as quasi-social agents. Building on this work, we investigate how students' confidence in AI assistance, perceived social expectations, and enjoyment of interaction relate to their intentions to use conversational AI, both directly and through perceptions of usefulness and ease of interaction.

To study these relationships, we extend TAM by incorporating trust, subjective norms, and perceived enjoyment alongside perceived usefulness and perceived ease of use. The proposed direct and indirect relationships among these constructs are examined using partial least squares structural equation modeling. The findings suggest that students' intentions to use conversational AI are closely associated with whether they find its assistance useful and its interactions enjoyable, while the perceived effort of using these tools does not independently predict their intentions. Students' trust in AI assistance and the social influence of others are also associated with students' behavioral intentions through how they evaluate the usefulness and interaction experience of a conversational AI tool. For educators and developers, these findings highlight the importance of understanding what students consider useful AI assistance, including whether it supports their understanding of course material or helps them complete academic tasks. This distinction matters when integrating AI into educational settings, as students may value assistance that simplifies their work without necessarily engaging with the reasoning behind it. Understanding these motivations can help educators establish expectations for AI use and guide research into how students' acceptance of pedagogically-aligned tools relates to their learning behaviors.

The remainder of this paper is organized as follows. We first review the theoretical foundations of technology acceptance models and examine the constructs of trust, subjective norms, and perceived enjoyment. Next, we present our theoretical framework and hypotheses, describing how these constructs are expected to relate to students' behavioral intentions to use a chatbot as a learning assistant. We then describe the methodology, including the study design, participant recruitment, survey measures, and the structural modeling approach used for analysis. Finally, we report the measurement and structural model results and discuss implications, limitations, and directions for future research.

\section{Theoretical Background \& Hypothesis Development}

The Technology Acceptance Model (TAM), developed by Davis \cite{davis1989technology}, has been a foundational framework for understanding individual technology adoption behaviors. TAM posits that an individual's intention to use a technology is primarily determined by two beliefs: perceived usefulness (PU), defined as the degree to which a person believes that using a particular system will enhance their performance, and perceived ease of use (PEOU), defined as the degree to which a person believes that using a system will be free from effort. These beliefs mediate the influence of external variables on behavioral intention, with perceived ease of use also influencing perceived usefulness.

In educational contexts, TAM has been commonly applied to understand student adoption of various technologies, including mobile learning applications \cite{park2012university} and other e-learning platforms \cite{abdullah2016developing, salloum2019exploring, pitts2025lessons}. Across contexts, perceived usefulness and perceived ease of use have frequently been associated with students' intentions to adopt educational technologies. For example, prior research on student adoption of e-portfolios and online learning systems has shown that when students perceive a system as useful for their academic performance and easy to navigate, they are more likely to integrate it into their learning routines \cite{abdullah2016investigating}.

Based on the original TAM structure, this study proposes the following hypotheses:

\begin{enumerate}[label=\textbf{H\arabic*:}]
    \item Perceived usefulness (PU) positively influences students' behavioral intention to use AI chatbots in education (BI).
    \item Perceived ease of use (PEOU) positively influences students' behavioral intention to use AI chatbots in education (BI).
    \item Perceived ease of use (PEOU) positively influences perceived usefulness of AI chatbots in education (PU).
\end{enumerate}

However, the TAM framework has demonstrated limitations when applied to emerging technologies such as AI. Recent studies have highlighted that TAM's approach may overlook important contextual factors that influence technology adoption decisions beyond the original model's scope; such as subjective/social norms \cite{abdullah2016developing}, facilitating conditions \cite{bamansoor2018adoption}, and relevant technical training \cite{bamansoor2018adoption}, which have been identified as significant external predictors of technology adoption in prior technology adoption research. Further, more recent AI acceptance research indicates that expectations and pre-use attitudes can predict post-use acceptance even when measured system performance primarily affects experience rather than acceptance directly \cite{ebermann2026user}. These limitations become apparent when considering AI chatbots, which generate responses through natural language and adapt explanations and feedback to learners, simulating conversational exchanges that resemble human tutoring \cite{pitts2025understanding}. This interactive capability distinguishes them from earlier systems with fixed or rule-based functions and leads us to hypothesize that students' adoption decisions are influenced by additional psychological and social factors beyond those outlined in TAM \cite{davis1989technology}.

To address these limitations, as shown in Figure 1, we extend TAM by incorporating three additional factors: trust, subjective norms, and perceived enjoyment. Trust reflects the extent to which students feel confident in AI-generated outputs despite the technology's “black box” nature \cite{von2021transparency, pitts2025understanding}. Subjective norms emphasize the influence of peers, instructors, and institutions in legitimizing or discouraging adoption \cite{schepers2007meta, ajzen1991theory}, while perceived enjoyment captures the motivational value of interacting with AI in a conversational and engaging way \cite{davis1992extrinsic,pitts2024proposed}.

\subsection{Extended Factors for AI in Education}
\subsubsection{Trust in AI}

Student-AI trust has been defined as “a student's willingness to rely on an AI system's guidance, feedback, and information under conditions of uncertainty and potential vulnerability” \cite{pitts2025understanding}. In technological contexts, trust is generally understood to encompass perceptions of system functionality (belief in the system's ability to function effectively), reliability (belief that the system acts consistently), and helpfulness (belief in the system's ability to assist effectively with a given task) \cite{mcknight2011trust}. Although, recent work suggests that trust in AI systems may require additional consideration due to the generative and probabilistic nature of AI outputs \cite{pitts2025understanding}.

Unlike educational technologies with predictable, rule-based functions, AI chatbots generate responses that are difficult for students to anticipate or independently verify. The potential for inaccurate, biased, or misleading outputs introduces uncertainty and risk for learners \cite{pitts2025understanding}. Moreover, the “black box” nature of many AI systems makes it challenging for students to understand how responses are produced, which can further amplify uncertainty regarding system reliability and competence \cite{von2021transparency}. In educational settings, this uncertainty places students in a position where they must decide whether to rely on AI-generated feedback, disengage from the tool, or engage cautiously while critically evaluating its outputs. Consequently, trust, and more specifically, the calibration of trust, has heightened importance in shaping how students engage with and benefit from AI tools \cite{pitts2025understanding}.

Prior research highlights that trust influences both initial adoption and continued use of technologies, particularly in systems characterized by uncertainty or risk \cite{albayati2024investigating,nazaretsky2025critical,salloum2018factors, polyportis2025understanding, choung2023trust, pitts2025understanding}. In the context of AI for education, appropriate levels of trust can motivate students to meaningfully engage with system-generated feedback and integrate AI guidance into their learning practices. Trust may also influence how students evaluate the interaction itself, as greater confidence in generated responses can increase their perceptions of the system's usefulness and reduce the effort associated with questioning or verifying its outputs. Conversely, insufficient trust may lead students to underutilize potentially helpful tools, while excessive or misplaced trust may result in overreliance on inaccurate outputs \cite{lyu2025understanding, ranalli2021l2, pitts2025reliance,pitts2026trust}. Building on this, we extend TAM to incorporate trust as a construct expected to influence students' perceptions of usefulness and ease of use, as well as their behavioral intention to adopt AI chatbots for learning. Accordingly, the following hypotheses are proposed:

\begin{enumerate}[label=\textbf{H\arabic*:}, start=4]
    \item Trust in AI systems positively influences students' behavioral intention to use AI chatbots in education (BI).
    \item Trust in AI systems positively influences perceived usefulness of AI chatbots in education (PU).
    \item Trust in AI systems positively influences perceived ease of use of AI chatbots in education (PEOU).
\end{enumerate}

\subsubsection{Subjective Norms and AI in Education}

Subjective norms, or social norms, refer to an individual's perception of social pressure to perform or refrain from a particular behavior \cite{ajzen1991theory}. In the context of technology adoption, subjective norms reflect the expectations of others such as peers, instructors, and institutional authorities. These social influences may be especially salient for AI adoption in academic settings, where norms surrounding appropriate AI use are still emerging and often contested.

In education more broadly, peer attitudes have been found to significantly influence student perceptions, as students often look to their classmates for cues regarding acceptable academic practices and technology use \cite{ballesteros2025influence}. Word-of-mouth peer influence, in particular, has demonstrated strong effects on behavioral intention, sometimes exceeding the influence of broader social norms \cite{ballesteros2025influence}. Instructors and institutions also play an important role in shaping students' learning behaviors. When instructors model responsible and effective AI use, for example, by providing clear guidelines and expectations for AI use in the classroom, providing AI-supported feedback in a responsible manner, or recommending AI study aids, students are more likely to view the technology as legitimate and beneficial \cite{huefactors}. Similarly, institutional support in the form of infrastructure, technical assistance, and clear usage guidelines has been shown to strengthen perceived usefulness and perceived ease of use for educational technologies \cite{sanchez2010motivational, polyportis2025understanding}. Beyond general support, institutional policy may also influence whether socially supported intentions become actual use. In a structural model of Dutch students' AI acceptance, Polyportis and Pahos \cite{polyportis2025understanding} found that institutional policy moderated the intention to use behavior relationship, and social influence had a direct significant influence. Strong facilitating conditions at the institutional level can therefore reinforce positive social norms around adoption \cite{sanchez2010motivational}, while environments lacking such support or where AI raises strong concerns about academic integrity may generate negative social pressures that discourage adoption despite perceived benefits \cite{ballesteros2025influence}. 

These social signals may also influence how students evaluate their interactions with AI. When AI use is accepted or encouraged by peers, instructors, or institutions, students may be more likely to view the technology as legitimate and trustworthy. Positive social expectations may also contribute to a more favorable experience with the technology, influencing how enjoyable students expect AI-supported learning to be. Subjective norms are therefore expected to relate to students' behavioral intentions both directly and indirectly through their perceptions of usefulness, ease of use, trust, and enjoyment. Hence, we propose the following hypotheses:

\begin{enumerate}[label=\textbf{H\arabic*:}, start=7]
    \item Subjective norms (SN) positively influence students' behavioral intention to use AI chatbots in education (BI).
    \item Subjective norms (SN) positively influence perceived usefulness of AI chatbots in education (PU).
    \item Subjective norms (SN) positively influence perceived ease of use of AI chatbots in education (PEOU).
    \item Subjective norms (SN) positively influence trust in AI chatbots in education.
    \item Subjective norms (SN) positively influence perceived enjoyment of AI chatbots in education (PE).
\end{enumerate}

\begin{figure*}[!t]
    \centering
    \includegraphics[width=0.7\textwidth]{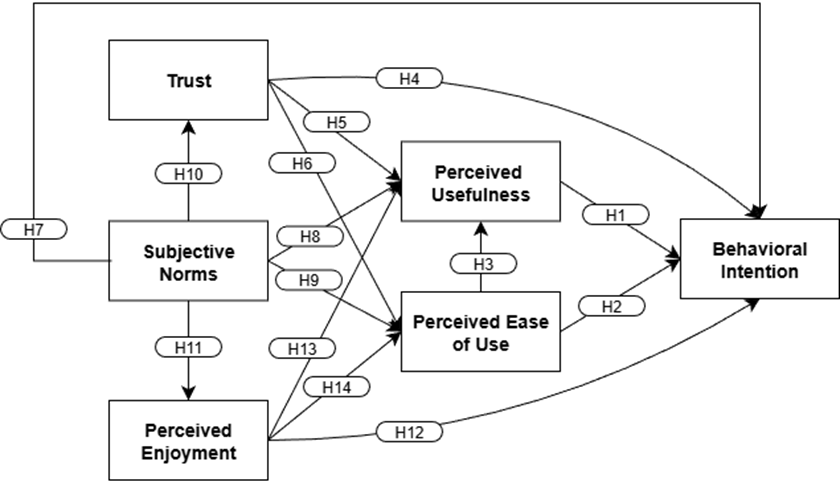}
    \caption{Proposed theoretical framework and hypotheses for students' adoption of conversational AI chatbots in education.}
    \label{fig:hypo-model}
\end{figure*}

\subsubsection{Perceived Enjoyment from Interactions with AI}

Perceived enjoyment refers to the extent to which using a technology is experienced as enjoyable in itself, independent of any anticipated performance or learning benefits \cite{davis1992extrinsic}. Enjoyment has long been recognized as a driver of voluntary technology adoption, particularly in educational settings where intrinsic motivation can significantly influence persistence and learning outcomes. When technologies are engaging and pleasant to use, students are more likely to participate actively, sustain their engagement over time, and explore the system beyond mandatory requirements. Prior studies on gamified platforms and interactive learning tools have consistently shown that perceived enjoyment enhances adoption and leads to deeper engagement with learning activities \cite{hamari2016challenging}. 

In the context of AI chatbots, enjoyment may stem from their conversational and interactive design. Unlike static e-learning systems, chatbots provide natural dialogue and personalized responses that can make learning feel more dynamic and supportive. Their ability to simulate conversational partners often triggers anthropomorphic responses, with students addressing them politely, attributing human-like qualities, and perceiving them as social actors \cite{nass1999people, reeves1996media}. This effect is consistent with the Computers as Social Actors (CASA) paradigm \cite{reeves1996media}, which suggests that people naturally apply social norms to their interactions with computers when these systems exhibit human-like cues. Research on chatbot use also indicates that users often experience them as friendly companions who not only provide information but also offer comfort, encouragement, and efficiency in completing tasks \cite{brandtzaeg2022my}. These experiences can generate positive affect and reduce the cognitive effort associated with learning, thereby reinforcing both engagement and adoption.

Perceived enjoyment may have a distinct influence on the adoption of AI chatbots relative to other constructs in TAM. Whereas perceived usefulness and ease of use emphasize instrumental and functional value, enjoyment captures the hedonic and affective dimensions of learning. In educational contexts, it is plausible that students who experience chatbot interactions as engaging may also perceive them as more useful and less effortful to use. Thus, enjoyment may represent both a motivational driver and a reinforcing mechanism for engagement with AI tools. Based on this reasoning, and as shown in Figure \ref{fig:hypo-model}, the following hypotheses are proposed: 
\begin{enumerate}[label=\textbf{H\arabic*:}, start=12]
    \item Perceived enjoyment (PE) positively influences students' behavioral intention to use AI chatbots in education (BI).
    \item Perceived enjoyment (PE) positively influences perceived usefulness of AI chatbots in education (PU).
    \item Perceived enjoyment (PE) positively influences perceived ease of use of AI chatbots in education (PEOU).
\end{enumerate} 

In addition to evaluating the proposed direct relationships, we examine how trust, subjective norms, and perceived enjoyment are associated with behavioral intention through other acceptance beliefs. Considering these indirect associations allows us to examine how students' confidence in AI assistance, social expectations, and interaction experiences relate to their evaluations of usefulness and ease of use, providing a more detailed account of their intentions to use conversational AI.

In the following section, we describe the study design, survey measures, and procedures used to evaluate the proposed theoretical framework and the hypothesized relationships among constructs.

\section{Methodology}

\subsection{Development of Instrument Measures}

A survey instrument was developed to assess the theoretical framework presented in Figure \ref{fig:hypo-model}. The instrument measured six constructs: behavioral intention (BI), perceived ease of use (PEOU), perceived usefulness (PU), trust, subjective norms (SN), and perceived enjoyment (PE). Each construct was assessed using four items adapted from previously established instruments \cite{venkatesh2000theoretical, pitts2024proposed, gulati2019design, Taylor1995-xe}. Responses were recorded on a five-point Likert scale ranging from 1 (Strongly Disagree) to 5 (Strongly Agree). The complete measurement items and their sources are provided in Appendix A.

\subsection{Procedures and Participants}

To evaluate the proposed framework (Figure~\ref{fig:hypo-model}), we used survey responses collected from undergraduate students at a large public research university through Qualtrics. The study was reviewed by the university's Institutional Review Board and classified as exempt research. Before participating, students reviewed an informed consent document and provided their consent. Participants completed questions about their demographic characteristics and academic background, including age, gender, grade-point average (GPA), major, year in university, and previous experience with AI. They also viewed a brief introductory video presenting AI chatbots and an example interaction. For the present analysis, 24 Likert-scale items corresponding to the six constructs in the proposed framework were used, presented in randomized order (Appendix A).

The survey received 293 responses. Of the 293 survey responses collected, 229 were retained for analysis after screening for completeness. This sample satisfied the commonly used 10-times guideline for PLS-SEM sample sizes \cite{Hair01042011}, which considers the larger of the maximum number of formative indicators measuring a construct or the maximum number of structural paths directed toward a construct. Participants had a mean age of 20.27 years, with 49.1\% identifying as female, 48.1\% as male, and 2.8\% preferred not to disclose or to self-describe gender. Most respondents (90.32\%) reported previous experience with AI chatbots, with an average experience rating of 4.44 out of 5. Regarding usage frequency, 63.31\% reported high-frequency use (daily or weekly), 31.00\% reported low-frequency use (monthly or less often), and 5.68\% reported no use of AI chatbots. 

\subsection{Data Analysis}

We analyzed the data using partial least squares structural equation modeling (PLS-SEM) in SmartPLS 4.0 \cite{ringle2024smartpls}. Following \cite{Hair01042011}, we used a two-stage approach: (1) evaluation of the measurement model and (2) evaluation of the structural model. This approach allowed us to assess the reliability and validity of the constructs before examining their hypothesized relationships.

For the measurement model, internal consistency reliability was assessed using Cronbach's alpha and composite reliability (CR), with values above 0.70 treated as satisfactory \cite{hair2019when}. Convergent validity was evaluated using indicator loadings and average variance extracted (AVE), with an AVE threshold of 0.50 \cite{hair2019when}. Discriminant validity was examined using the Fornell--Larcker criterion (comparing the square root of AVE to inter-construct correlations), cross-loadings \cite{fornell1981evaluating, Hair01042011}, and the heterotrait–monotrait (HTMT) ratio with bootstrapped 95\% confidence intervals \cite{henseler2015new}. Further, multicollinearity among predictor constructs was assessed using inner variance inflation factors (VIFs), with values below 5.0 indicating acceptable levels of shared variance \cite{Hair01042011}.

For the structural model, we examined standardized path coefficients, their statistical significance, and the explanatory power of the endogenous constructs. Consistent with prior work, standardized path coefficients of 0.05, 0.10, and 0.25 were interpreted as small, moderate, and large effects, respectively \cite{keith2019multiple}. Statistical significance of path coefficients was evaluated via bootstrapping with 5{,}000 resamples \cite{banjanovic2016confidence}. In addition to the hypothesized direct relationships, we examined total indirect effects to assess how trust, subjective norms, and perceived enjoyment were associated with behavioral intention through other constructs in the model. Effect sizes ($f^2$) were calculated to determine the contribution of each predictor to its respective endogenous variable, with values of 0.02, 0.15, and 0.35 representing small, medium, and large effects, respectively \cite{cohen1988set}. Model fit was evaluated using the standardized root mean square residual (SRMR) for both the saturated and estimated models. SRMR values below 0.08 are generally considered acceptable indicators of overall model fit \cite{hu1999cutoff}. Model explanatory power was evaluated using $R^2$ values for each endogenous construct, interpreted according to Chin's \cite{chin1998partial} guidelines (0.19 = weak, 0.33 = moderate, 0.67 = substantial). 

\section{Results}
This section reports the PLS-SEM results in two stages. First, we assess the measurement model to examine discriminant validity and internal consistency. The structural model (Figure 1) is then evaluated, testing the hypothesized relationships and determining how much variance the model explains.

\subsection{Assessment of Measurement Model}

We assessed the measurement model by examining the internal consistency and convergent validity of the six constructs in the proposed framework. Reliability was evaluated using Cronbach's alpha and composite reliability (CR), which ranged from 0.846 to 0.947 and 0.895 to 0.961, respectively. All values exceeded the recommended threshold of 0.70, indicating satisfactory internal consistency across the measurement scales \cite{hair2019when}. To evaluate convergent validity, we examined individual factor loadings and average variance extracted (AVE). All indicators except PEOU3 had factor loadings above 0.70. Although PEOU3 fell below this threshold, it was retained because the perceived ease of use construct continued to satisfy the recommended CR and AVE criteria. Across the six constructs, AVE values ranged from 0.685 to 0.862, exceeding the recommended minimum of 0.50 \cite{hair2019when}. These results support the reliability and convergent validity of the measurement model (Appendix B).

\begin{table}[t]
\centering
\caption{Fornell--Larcker matrix. Diagonal entries (bold) are the square roots of AVE; off-diagonal entries are construct correlations.}
\setlength{\tabcolsep}{6pt}
\renewcommand{\arraystretch}{1.15}
\begin{tabular}{lcccccc}
\toprule
 & BI & PU & PEOU & Trust & SN & PE \\
\midrule
BI    & \textbf{0.928} & 0.799 & 0.672 & 0.715 & 0.577 & 0.733 \\
PU    & 0.799 & \textbf{0.884} & 0.746 & 0.827 & 0.615 & 0.758 \\
PEOU  & 0.672 & 0.746 & \textbf{0.828} & 0.628 & 0.451 & 0.623 \\
Trust & 0.715 & 0.827 & 0.628 & \textbf{0.900} & 0.582 & 0.716 \\
SN    & 0.577 & 0.615 & 0.451 & 0.582 & \textbf{0.885} & 0.560 \\
PE    & 0.733 & 0.758 & 0.623 & 0.716 & 0.560 & \textbf{0.893} \\
\bottomrule
\end{tabular}
\label{tab:matrix}
\end{table}

\begin{table}[t]
\centering
\caption{HTMT ratios and 95\% confidence intervals.}
\label{tab:htmt}
\begin{tabular}{lcc}
\toprule
Construct Pair & HTMT & 95\% CI \\
\midrule
PE -- BI       & 0.772 & [0.692, 0.840] \\
PEOU -- BI     & 0.715 & [0.611, 0.805] \\
PEOU -- PE     & 0.663 & [0.560, 0.751] \\
PU -- BI       & 0.858 & [0.794, 0.911] \\
PU -- PE       & 0.819 & [0.753, 0.875] \\
PU -- PEOU     & 0.825 & [0.744, 0.898] \\
SN -- BI       & 0.615 & [0.490, 0.719] \\
SN -- PE       & 0.602 & [0.466, 0.721] \\
SN -- PEOU     & 0.499 & [0.364, 0.623] \\
SN -- PU       & 0.676 & [0.569, 0.769] \\
Trust -- BI    & 0.761 & [0.678, 0.832] \\
Trust -- PE    & 0.766 & [0.679, 0.842] \\
Trust -- PEOU  & 0.681 & [0.577, 0.774] \\
Trust -- PU    & 0.904 & [0.850, 0.948] \\
Trust -- SN    & 0.630 & [0.512, 0.732] \\
\bottomrule
\end{tabular}
\end{table}

\subsubsection{Discriminant Validity and Construct Correlations}

We examined discriminant validity through the Fornell–Larcker criterion and heterotrait–monotrait (HTMT) ratios. Convergent validity was first confirmed, which serves as a prerequisite \cite{bagozzi1982representing}. As shown in Table \ref{tab:matrix}, through the Fornell–Larcker criterion, the square root of AVE for each construct (shown in bold on the diagonal) exceeded its correlations with other constructs, confirming discriminant validity. Construct correlations ranged from moderate (0.451 between PEOU and SN) to high (0.799 between PU and BI). 

We further assessed discriminant validity using HTMT ratios and their bootstrapped 95\% confidence intervals (Table 2). All construct pairs except trust and PU had HTMT values below 0.90. The trust and PU pair had an HTMT ratio of 0.904, with a confidence interval of [0.850, 0.948] that excluded 1.00. Although the results generally support discriminant validity, the elevated ratio between trust and perceived usefulness suggests some conceptual and statistical overlap between these constructs. This relationship is discussed further in Section 5.

\begin{table}[t]
\centering
\caption{Inner VIF values with 95\% confidence intervals.}
\label{tab:vif}
\begin{tabular}{lcc}
\toprule
Path & VIF & 95\% CI \\
\midrule
PE $\rightarrow$ BI       & 2.695 & [2.167, 3.365] \\
PE $\rightarrow$ PEOU     & 2.244 & [1.841, 2.771] \\
PE $\rightarrow$ PU       & 2.244 & [1.841, 2.771] \\
PEOU $\rightarrow$ BI     & 2.372 & [1.921, 2.954] \\
PU $\rightarrow$ BI       & 5.201 & [4.042, 6.601] \\
SN $\rightarrow$ BI       & 1.749 & [1.435, 2.149] \\
SN $\rightarrow$ PE       & 1.000 & [1.000, 1.000] \\
SN $\rightarrow$ PEOU     & 1.650 & [1.359, 2.041] \\
SN $\rightarrow$ PU       & 1.650 & [1.359, 2.041] \\
SN $\rightarrow$ Trust    & 1.000 & [1.000, 1.000] \\
Trust $\rightarrow$ BI    & 3.562 & [2.722, 4.671] \\
Trust $\rightarrow$ PEOU  & 2.331 & [1.820, 3.001] \\
Trust $\rightarrow$ PU    & 2.331 & [1.820, 3.001] \\
\bottomrule
\end{tabular}
\end{table}

\begin{table}[t]
\centering
\caption{Hypothesis testing results.}
\label{tab:hypotheses}
\begin{tabular}{clccc}
\toprule
Hyp. & Path & $\beta$ & $p$ value & Result \\
\midrule
H1  & PU $\rightarrow$ BI       & 0.385 & $< .001$ & Supported \\
H2  & PEOU $\rightarrow$ BI     & 0.142 & .069     & Not supported \\
H3  & PEOU $\rightarrow$ PU     & 0.301 & $< .001$ & Supported \\
H4  & Trust $\rightarrow$ BI    & 0.077 & .288     & Not supported \\
H5  & Trust $\rightarrow$ PU    & 0.428 & $< .001$ & Supported \\
H6  & Trust $\rightarrow$ PEOU  & 0.354 & $< .001$ & Supported \\
H7  & SN $\rightarrow$ BI       & 0.093 & .116     & Not supported \\
H8  & SN $\rightarrow$ PU       & 0.121 & .005     & Supported \\
H9  & SN $\rightarrow$ PEOU     & 0.055 & .462     & Not supported \\
H10 & SN $\rightarrow$ Trust    & 0.580 & $< .001$ & Supported \\
H11 & SN $\rightarrow$ PE       & 0.559 & $< .001$ & Supported \\
H12 & PE $\rightarrow$ BI       & 0.247 & .002     & Supported \\
H13 & PE $\rightarrow$ PU       & 0.197 & $< .001$ & Supported \\
H14 & PE $\rightarrow$ PEOU     & 0.336 & $< .001$ & Supported \\
\bottomrule
\end{tabular}
\end{table}

\begin{table}[ht]
\centering
\caption{$f^2$ Effect Sizes}
\label{tab:f2}
\begin{tabular}{lcccc}
\toprule
Path & $f^2$ & SD & $t$ & $p$ \\
\midrule
PE $\rightarrow$ BI       & 0.084 & 0.053 & 1.430 & 0.153 \\
PE $\rightarrow$ PEOU     & 0.105 & 0.051 & 1.875 & 0.061 \\
PE $\rightarrow$ PU       & 0.172 & 0.060 & 2.731 & 0.006 \\
PEOU $\rightarrow$ BI     & 0.039 & 0.036 & 0.850 & 0.396 \\
PU $\rightarrow$ BI       & 0.108 & 0.059 & 1.606 & 0.108 \\
SN $\rightarrow$ BI       & 0.024 & 0.023 & 0.714 & 0.475 \\
SN $\rightarrow$ PE       & 0.477 & 0.146 & 3.124 & 0.002 \\
SN $\rightarrow$ PEOU     & 0.010 & 0.013 & 0.254 & 0.799 \\
SN $\rightarrow$ PU       & 0.053 & 0.036 & 1.305 & 0.192 \\
SN $\rightarrow$ Trust    & 0.529 & 0.143 & 3.551 & $< .001$ \\
Trust $\rightarrow$ BI    & 0.010 & 0.012 & 0.454 & 0.650 \\
Trust $\rightarrow$ PEOU  & 0.111 & 0.058 & 1.771 & 0.077 \\
Trust $\rightarrow$ PU    & 0.526 & 0.153 & 3.341 & 0.001 \\
\bottomrule
\end{tabular}
\end{table}

\begin{table}[ht]
\centering
\caption{Total Indirect Effects}
\label{tab:path_coefficients}
\begin{tabular}{lcccc}
\toprule
\textbf{Path} & $\beta$ & SD & $t$ & $p$ \\
\midrule
PE $\rightarrow$ BI    & 0.163 & 0.040 & 4.090  & $< .001$ \\
SN $\rightarrow$ BI    & 0.483 & 0.046 & 10.538 & $< .001$ \\
SN $\rightarrow$ PEOU  & 0.395 & 0.053 & 7.383  & $< .001$ \\
SN $\rightarrow$ PU    & 0.477 & 0.049 & 9.759  & $< .001$ \\
Trust $\rightarrow$ BI & 0.256 & 0.050 & 5.086  & $< .001$ \\
\bottomrule
\end{tabular}
\end{table}

\subsection{Assessment of Structural Model}

\begin{figure*}[!t]
    \centering
    \includegraphics[width=0.7\textwidth]{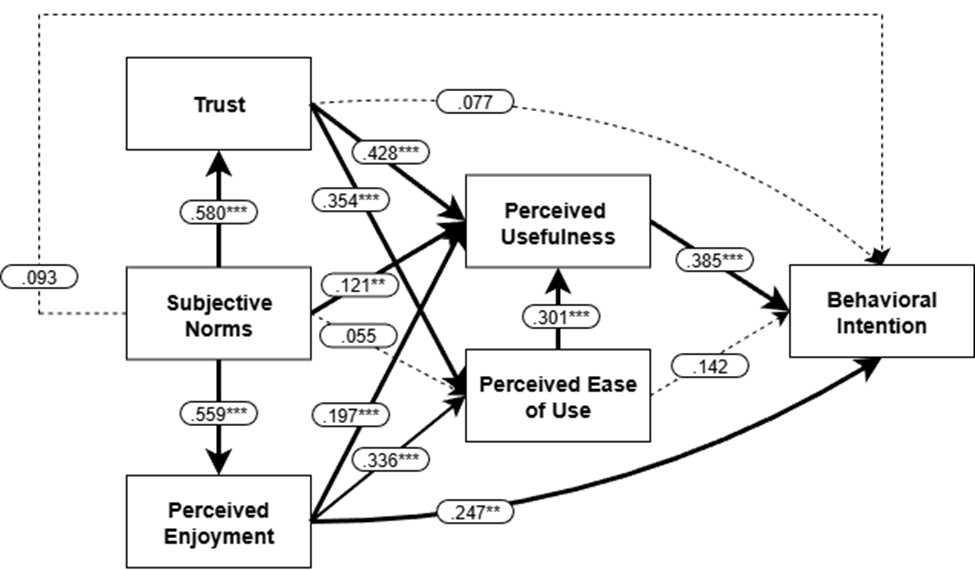}
    \caption{PLS-SEM structural model results with standardized path coefficients; solid lines denote significant paths and dashed lines denote non-significant paths (*p < .05, **p < .01, ***p < .001).}
    \label{fig:structuralmodel}
\end{figure*}

After confirming the measurement model's validity, we examined the structural relationships among constructs. We first assessed inner VIF values to evaluate collinearity among the predictor constructs. As shown in Table~\ref{tab:vif}, the reported VIF values ranged from 1.000 to 5.201. All reported values were below the commonly used threshold of 5.0 except for the path from PU to BI, which had a VIF of 5.201. This value exceeded the threshold, indicating some collinearity among the predictors of BI, while the remaining portions of the model did not exhibit elevated VIF values. The PU $\rightarrow$ BI coefficient should therefore be interpreted alongside the construct correlations and its $f^2$ effect size.

The structural model demonstrated substantial explanatory power for BI and PU, explaining 70.2\% of the variance in BI ($R^{2}=0.702$) and 80.5\% of the variance in PU ($R^{2}=0.805$). The model also explained 46.6\% of the variance in PEOU ($R^{2}=0.466$), 34.2\% of the variance in Trust ($R^{2}=0.342$), and 31.7\% of the variance in PE ($R^{2}=0.317$). Based on Chin's guidelines \cite{chin1998partial}, these values indicate substantial explanatory power for BI and PU, moderate explanatory power for PEOU and Trust, and explanatory power approaching the moderate threshold for PE.

Analysis of path coefficients revealed significant relationships for 10 out of 14 hypothesized paths (See Table~\ref{tab:hypotheses} and Figure~\ref{fig:structuralmodel}). Users' intention to use an AI learning assistant was strongly influenced by PU ($\beta = 0.385, p < 0.001$) and PE ($\beta = 0.247, p = 0.002$), but not significantly by PEOU ($\beta = 0.142, p = 0.069$), Trust ($\beta = 0.077, p = 0.288$), or SN ($\beta = 0.093, p = 0.116$). PEOU significantly influenced PU ($\beta = 0.301, p < 0.001$). Trust strongly influenced both PEOU ($\beta = 0.354, p < 0.001$) and PU ($\beta = 0.428, p < 0.001$). PE significantly influenced BI ($\beta = 0.247, p = 0.002$), PU ($\beta = 0.197, p < 0.001$), and PEOU ($\beta = 0.336, p < 0.001$). SN significantly influenced Trust ($\beta = 0.580, p < 0.001$), PE ($\beta = 0.559, p < 0.001$), and PU ($\beta = 0.121, p = 0.005$), but showed non-significant effects on PEOU ($\beta = 0.055, p = 0.462$). 

The reported $f^2$ values provide additional context for the relative contribution of each predictor. Large effects were observed for SN predicting Trust ($f^2=0.529$), Trust predicting PU ($f^2=0.526$), and SN predicting PE ($f^2=0.477$). PE had a medium effect on PU ($f^2=0.172$). The remaining reported effects were small or negligible, ranging from 0.010 to 0.111. Although PU had the largest direct path coefficient predicting BI, its reported $f^2$ value was 0.108, indicating a smaller unique contribution after accounting for the other predictors of BI. This result is also consistent with the elevated VIF for the hypothesized path from PU to BI.

The indirect-effect results clarify how Trust, SN, and PE are associated with the endogenous constructs. PE had a significant total indirect effect on BI ($\beta = 0.163, p < 0.001$), while Trust also had a significant total indirect effect on BI ($\beta = 0.256, p < 0.001$). SN had significant total indirect effects on BI ($\beta = 0.483, p < 0.001$), PEOU ($\beta = 0.395, p < 0.001$), and PU ($\beta = 0.477, p < 0.001$). These findings indicate that Trust and SN remained associated with behavioral intention despite their non-significant direct paths, primarily through their relationships with usefulness, ease of use, and enjoyment. The SRMR for the saturated model was 0.070, below the recommended upper threshold of 0.08 and indicating acceptable model fit.

\section{Discussion}

This study examines factors associated with students' intentions to use conversational AI chatbots in educational settings using a technology acceptance framework. The model explained 70.2\% of the variance in behavioral intention and 80.5\% of the variance in perceived usefulness, indicating that the extended framework accounted for a substantial portion of students' acceptance judgments. These findings suggest that students' intentions to use conversational AI reflect interconnected judgments about its usefulness, the experience of interacting with it, confidence in its outputs, and the social context surrounding its use. Specifically, perceived usefulness had the strongest direct association with behavioral intention, followed by perceived enjoyment, while perceived ease of use was associated with usefulness but did not have a significant direct association with intention. Perceived enjoyment was associated with behavioral intention both directly and indirectly, while trust and subjective norms had significant total indirect associations with intention through other acceptance beliefs.

\subsection{Perceived Usefulness as the Primary Determinant of Usage Intentions}

Perceived usefulness had the strongest direct association with students' behavioral intention to use conversational AI chatbots. The perceived usefulness items represent several related forms of academic value. PU1 and PU4 assess whether students view an AI chatbot as generally useful while learning, PU3 focuses on whether it provides valuable academic support when needed, and PU2 assesses whether it decreases frustration during learning. Students were therefore more likely to intend to use an AI chatbot when they believed that it could provide accessible support and improve their experience of completing academic work. This finding aligns with prior research demonstrating the importance of perceived usefulness in the acceptance of educational technologies, mobile learning systems, and online platforms \cite{abdullah2016investigating, salloum2018factors, pitts2025lessons}. 

Perceived usefulness also served as a ``hub'' construct in the model, as trust, perceived enjoyment, subjective norms, and perceived ease of use were all positively associated with it. In turn, perceived usefulness had the strongest direct association with behavioral intention, suggesting that students' evaluations of conversational AI's usefulness reflect several interconnected beliefs about the technology. For instance, a chatbot may be viewed as more useful when its responses appear dependable, when communication feels manageable, when the interaction is engaging, and when its use is viewed positively by important others. However, although perceived usefulness had the largest direct coefficient predicting behavioral intention ($\beta=0.385$), its $f^2$ value was 0.108, indicating a relatively small unique contribution after accounting for the other predictors. The VIF for the PU $\rightarrow$ BI path was 5.201, slightly exceeding the prespecified threshold of 5.0 and suggesting some collinearity among the predictors of intention. The high association between perceived usefulness and trust further indicates that students' evaluations of a chatbot's academic value and the dependability of its assistance are closely related. Perceived usefulness should therefore be interpreted as a strongest predictor in the model while recognizing its overlap with other acceptance beliefs.

Importantly, perceived usefulness reflects a broad judgment that a chatbot helps with academic work, but it does not distinguish what students mean by ``help.'' For some students, usefulness may reflect pedagogically supportive interactions, such as clearer explanations, feedback, or support for reasoning \cite{fan2023exploring, oli2024towards}. For others, usefulness may reflect task offloading, where the chatbot produces answers or reduces the work needed to complete an assignment. This distinction matters because when usefulness is interpreted mainly as convenience, students may be more prone to overreliance and less likely to scrutinize outputs \cite{pitts2025reliance, nazaretsky2025critical}. Future research should distinguish between different forms of perceived usefulness, such as assistance that supports conceptual understanding and assistance that simplifies task completion, and examine how these perceptions relate to AI use, reliance, and learning outcomes.

From a practical standpoint, these findings highlight the importance of considering what students value when using AI chatbots for academic work. For educators, this includes establishing expectations for how AI assistance should contribute to learning activities, such as using generated explanations to check understanding, evaluate reasoning, or revise one's own work. For developers, it suggests examining whether features such as structured explanations, verification prompts, and reasoning checks are perceived as useful by students and encourage meaningful engagement with course material. This approach leverages the same adoption pathway identified in our theoretical framework, while steering “usefulness” toward learning-oriented value instead of uses that primarily replace opportunities for students' critical thinking.

\subsection{Perceived Enjoyment as a Direct and Indirect Contributor}

Perceived enjoyment was the second significant direct predictor of behavioral intention ($\beta=0.247$) and also had a significant total indirect effect on intention ($\beta=0.163$). The enjoyment items capture several aspects of students' affective experience with conversational AI. PE1 assesses whether an AI chatbot would make learning more interesting, PE2 focuses on engagement, PE3 focuses on enjoyment during learning, and PE4 measures whether students would enjoy using the chatbot as a learning aid. These results indicate that students' intentions are associated with what the chatbot helps them accomplish and how enjoyable they expect the interaction to be.

Perceived enjoyment also significantly predicted perceived usefulness ($\beta=0.197$) and perceived ease of use ($\beta=0.336$). Its effect on perceived usefulness was medium in magnitude ($f^2=0.172$), while its reported effects on behavioral intention ($f^2=0.084$) and perceived ease of use ($f^2=0.105$) were small. This pattern suggests that enjoyment contributes to acceptance in several connected ways. Students may be more willing to use an AI chatbot because the interaction is engaging, while that positive experience may also lead them to judge the system as more academically useful and easier to interact with.

This finding is relevant to the design of conversational interfaces because their value is experienced through an ongoing exchange. Responsiveness, conversational pacing, language choice, opportunities to ask follow-up questions, and supportive feedback may influence how interaction feels engaging. However, enjoyment should remain connected to productive learning behaviors. A conversational system may be pleasant to use while still encouraging passive acceptance of responses or unnecessary dependence. Designers should therefore consider affective engagement alongside features that prompt evaluation, reflection, and active participation.

Additionally, subjective norms had a large effect on perceived enjoyment ($f^2=0.477$), indicating that enjoyment may also be socially situated. Students may experience an AI chatbot more positively when its use is accepted or encouraged by peers and other people important to them. Perceived enjoyment in this context may therefore reflect both interface qualities and the broader social meaning attached to interacting with AI for academic work.

\subsection{Perceived Ease of Use as an Indirect Contributor}
Perceived ease of use did not exert a significant direct effect on students' behavioral intention to use a conversational AI chatbot once other factors were considered. Instead, ease of use influenced adoption indirectly through its effect on perceived usefulness. This differs from many prior applications of TAM in educational contexts, where ease of use directly predicts usage intentions \cite{abdullah2016developing, salloum2018factors, pitts2025lessons}. 

One interpretation is that conversational AI systems reduce the salience of effort-related considerations by lowering interaction barriers. AI chatbots allow students to engage through familiar conversational exchanges, such as asking questions or requesting clarification, without requiring substantial learning of interface conventions or workflows. When interaction feels manageable, ease of use may function less as an independent motivator and more as an enabling condition that allows students to access and assess the system's potential value. This indirect role points to how, in broader conversational AI applications such as embodied social robots, voice-based assistants, and dialogue-driven learning agents, usability may primarily enable interaction rather than motivate adoption in its own right. Under these conditions, users' adoption decisions are likely influenced less by interaction mechanics and more by how system outputs are interpreted, trusted, and experienced. Factors such as confidence in system responses, affective engagement during interaction, and cues from the surrounding social environment likely become more influential as interaction barriers and usability concerns decrease. This framing helps explain why perceived ease of use remains linked to perceived usefulness while exerting less direct influence on behavioral intention in the context of this study, and it motivates closer consideration of trust in students' interactions with AI.

\subsection{Trust as an Indirect Reinforcer of Usage Intentions}

Although trust did not have a significant direct relationship with students' behavioral intention to use a conversational AI chatbot, it had a significant total indirect effect on intention and strong relationships with both perceived usefulness and perceived ease of use. This pattern suggests that trust may contribute to acceptance primarily by influencing how students evaluate and experience AI-generated assistance, instead of functioning as an independent motivation for use. This role becomes clearer when considering how students approach AI-generated responses. When trust is low, students may interact cautiously by verifying outputs, limiting use to low-stakes tasks, or disengaging after encountering questionable responses \cite{pitts2025understanding, pitts2025reliance}. These behaviors increase the effort required to use the system and reduce the extent to which its outputs are perceived as helpful. When trust is higher, students are more likely to engage fluidly with AI chatbots and incorporate responses into their academic activities with less friction \cite{pitts2025understanding, pitts2025reliance}. In this way, trust affects adoption indirectly by influencing how system outputs are experienced during interaction. This differs from how trust has often been conceptualized in prior work involving educational technologies, where trust commonly relates to institutional reliability, system stability, or data security \cite{pitts2025understanding}. In the present study, the indirect role of trust reflects students' perceived uncertainty about the quality, appropriateness, and reliability of generated content from conversational AI tools \cite{pitts2025understanding}.

These results also carry implications for the development of pedagogically-aligned AI tools. Design features that make reliability and limitations visible, such as citations, confidence indicators, or optional explanations, may help students judge when and how to rely on chatbot outputs. Progressive disclosure approaches, where surface-level signals are available by default and deeper reasoning can be accessed on demand, offer one potential strategy.

\subsection{The Role of Subjective Norms in Students' AI Use}

Subjective norms did not have a significant direct relationship with behavioral intention, yet they had the largest total indirect effect on intention among the constructs examined in the model. Subjective norms were also strongly related to trust and perceived enjoyment and were significantly associated with perceived usefulness. This suggests that social influence may contribute to acceptance less through direct pressure to use AI and more through the beliefs and experiences students develop around its use. Students encounter AI within classroom and institutional environments where norms surrounding appropriate use are still emerging. Consistent with prior work on social influence in educational settings \cite{sanchez2010motivational, ballesteros2025influence}, when peers, instructors, or institutions signal that AI use is acceptable and supported, students are more likely to view AI chatbots as trustworthy, enjoyable, and useful. When norms are ambiguous or restrictive, students may discount the technology regardless of its technical capabilities. In this sense, subjective norms function as a legitimacy signal that frames whether AI tools are perceived as appropriate for learning.

This pattern also helps contextualize broader implications of adoption. To the extent that subjective norms relate to trust and enjoyment more strongly than to behavioral intention, institutional guidance and classroom practices may be especially influential for how students experience AI in education, not only whether they use such tools. When policies clearly articulate the purposes of AI in education, including where use is encouraged and where it may compromise academic integrity, students can rely less on informal peer norms to infer what is acceptable. In that sense, policies that specify appropriate learning contexts while drawing clear boundaries around misuse can provide practical structure for students' decisions. Recent reviews of AI governance point to a growing number of institutional and national guidelines \cite{correa2023worldwide}, and meta-syntheses of educational policy document how universities and broader educational institutions are beginning to codify AI's role in teaching and learning \cite{funa2025policy}. At the same time, these syntheses indicate that policy development is uneven and often too general to guide day-to-day classroom decisions, highlighting the need for clearer, context-sensitive guidance that supports consistent practice across courses and instructors.

\subsection{Limitations and Future Work}

Interpreting these findings requires consideration of the study's limitations, which in turn point toward future research directions. First, the survey design captured perceptions and intentions at a single point in time. The findings should therefore be understood as a snapshot of adoption reflecting one moment in the ongoing integration of AI in education, with patterns likely to change as new tools are introduced, others fade from use, and ongoing improvements address challenges such as transparency and reliability. Second, the study relied on self-reported measures of behavioral intention rather than actual usage behaviors. Although research has demonstrated strong correlations between intentions and behaviors in technology acceptance contexts \cite{venkatesh2003user}, future studies should examine actual AI chatbot usage patterns and their relationship to the factors identified in this model. Behavioral measures could include frequency of use, duration of interactions, types of tasks performed, and persistence of use over time, providing a more comprehensive understanding of student use of AI in education.

Additionally, the sample was drawn from a specific institutional context, which may limit the generalizability of findings to other educational settings with different technological infrastructures, cultural norms, or AI policies. The study also focused on general AI chatbot adoption rather than examining specific educational applications or use cases. Students' perceptions and adoption behaviors may vary depending on whether AI chatbots are used for homework assistance, research support, language learning, or other specific tasks. Future research should investigate how acceptance patterns differ across specific educational use cases and student demographics. Examining potential moderating factors such as AI literacy, academic discipline, or learning preferences could reveal when different factors become more or less important for adoption. As AI capabilities continue to advance, continuous monitoring of adoption patterns will help maintain the theoretical and practical relevance of human-factor models.

Finally, this study did not measure learning outcomes or the educational quality of students' AI use. A student may intend to use a chatbot and perceive it as useful without that use improving understanding, performance, or other learning outcomes. Future research should connect acceptance and usage measures with outcomes such as learning gains, task performance, quality of reasoning, reliance on AI-generated assistance, and retention. Doing so would help distinguish factors that encourage students to use conversational AI from those associated with educationally productive use.

\section{Conclusion}

This study used a technology acceptance framework to examine students' intention to use conversational AI in educational settings. Perceived usefulness was the strongest predictor of intention, while perceived ease of use related to intention mainly through usefulness. Trust, perceived enjoyment, and subjective norms contributed by influencing how students interpreted conversational AI interactions and evaluated the academic value of their outputs. The absence of a significant direct effect of perceived ease of use suggests that conversational interfaces may reduce the salience of usability concerns once basic interaction barriers are met. In this setting, specific to AI in education, adoption appears to be influenced less by effort-related considerations and more by how system outputs are understood, relied upon, and situated within classroom environments.

Theoretically, these findings support that established technology acceptance constructs remain relevant for understanding conversational AI use in education, while their relationships operate differently in ways that reflect conversational interaction and generative system behavior. In practice, the results point to the importance of transparent and reliable outputs, clear academic value, appropriately calibrated trust, and meaningful pedagogically-aligned interactions that sustain engagement. Aligning tool design and classroom integration with these priorities can support responsible engagement and encourage more deliberate, reflective, and motivated learning.







\setcounter{enumiv}{0}
\bibliographystyle{unsrt}
\bibliography{acceptance}

\clearpage

\appendix
\section{Instrument Measures}

\begin{table}[h]
\centering
\label{tab:instrument-measures}
\renewcommand{\arraystretch}{1.1}
\Large
\resizebox{\textwidth}{!}{%
\begin{tabular}{l l p{8cm} l}
\toprule
\textbf{Construct} & \textbf{Item} & \textbf{Question Text} & \textbf{Reference} \\
\midrule
\textbf{Behavioral Intention (BI)} & BI1 & I would use an AI chatbot while I am learning. & Adapted from Venkatesh \& Davis (2000) \\
 & BI2 & I plan to use an AI chatbot while learning in the future. & \\
 & BI3 & If given permission to use an AI chatbot for an academic course, I would use it. & \\
 & BI4 & If given permission to use an AI chatbot for an academic course, I predict that I would use it. & \\
\midrule
\textbf{Perceived Usefulness (PU)} & PU1 & An AI chatbot would be a useful tool to have while learning. & Adapted from Davis et al. (1989) \\
 & PU2 & An AI chatbot would decrease my frustration while learning. & \\
 & PU3 & An AI chatbot would provide valuable academic support when I require it. & \\
 & PU4 & An AI chatbot would be of use to me while I am learning. & \\
\midrule
\textbf{Perceived Ease of Use (PEOU)} & PEOU1 & Communicating with an AI chatbot is easy to do. & Adapted from Davis et al. (1989) \\
 & PEOU2 & I would find it easy to use an AI chatbot as a learning aid. & \\
 & PEOU3 & I would be able to easily understand an AI chatbot. & \\
 & PEOU4 & I find AI chatbots easy to communicate with. & \\
\midrule
\textbf{Trust} & Trust1 & I could depend on an AI chatbot for assistance while I am learning. & Adapted from Gulati et al. (2019) \\
 & Trust2 & I could rely on an AI chatbot for assistance while I am learning. & \\
 & Trust3 & In general, I could count on an AI chatbot for assistance while learning. & \\
 & Trust4 & I could trust an AI chatbot for assistance while I am learning. & \\
\midrule
\textbf{Subjective Norms (SN)} & SN1 & My peers would think I should use an AI chatbot as a learning aid. & Adapted from Taylor \& Todd (1995) \\
 & SN2 & In general, people who I like would think I should use an AI chatbot as a learning aid. & \\
 & SN3 & People who are important to me would think I should use an AI chatbot as a learning aid. & \\
 & SN4 & Individuals who are significant to me would think I should use an AI chatbot as a learning aid. & \\
\midrule
\textbf{Perceived Enjoyment (PE)} & PE1 & An AI chatbot would make learning more interesting. & Adapted from Pitts et al. (2024) \\
 & PE2 & An AI chatbot would make learning more engaging. & \\
 & PE3 & An AI chatbot would make learning more enjoyable. & \\
 & PE4 & I would enjoy using an AI chatbot as a learning aid. & \\
\bottomrule
\end{tabular}}
\end{table}

\clearpage
\section{Evaluation of Survey Instrument}
\begin{table}[h]
\centering
\label{tab:survey-eval}
\renewcommand{\arraystretch}{1.2}
\scriptsize 
\resizebox{\textwidth}{!}{%
\begin{tabular}{l l c c c c c}
\toprule
\textbf{Construct} & \textbf{Item} & \textbf{Factor Loading} & \textbf{M (SD)} & \textbf{Cronbach's $\alpha$} & \textbf{CR} & \textbf{AVE} \\
\midrule
\textbf{Behavioral Intention (BI)} & BI1 & .925 & 3.72 (1.13) & \multirow{4}{*}{.947} & \multirow{4}{*}{.961} & \multirow{4}{*}{.862} \\
 & BI2 & .919 & 3.73 (1.19) &  &  &  \\
 & BI3 & .930 & 3.91 (1.11) &  &  &  \\
 & BI4 & .940 & 3.85 (1.15) &  &  &  \\
\midrule
\textbf{Perceived Usefulness (PU)} & PU1 & .915 & 3.82 (1.08) & \multirow{4}{*}{.906} & \multirow{4}{*}{.934} & \multirow{4}{*}{.781} \\
 & PU2 & .830 & 3.46 (1.15) &  &  &  \\
 & PU3 & .882 & 3.63 (1.05) &  &  &  \\
 & PU4 & .905 & 3.82 (1.03) &  &  &  \\
\midrule
\textbf{Perceived Ease of Use (PEOU)} & PEOU1 & .875 & 3.80 (0.97) & \multirow{4}{*}{.846} & \multirow{4}{*}{.895} & \multirow{4}{*}{.685} \\
 & PEOU2 & .853 & 3.82 (0.99) &  &  &  \\
 & PEOU3 & .691 & 3.94 (0.91) &  &  &  \\
 & PEOU4 & .874 & 3.73 (1.05) &  &  &  \\
\midrule
\textbf{Trust} & Trust1 & .892 & 3.37 (1.16) & \multirow{4}{*}{.921} & \multirow{4}{*}{.944} & \multirow{4}{*}{.810} \\
 & Trust2 & .917 & 3.45 (1.13) &  &  &  \\
 & Trust3 & .914 & 3.48 (1.18) &  &  &  \\
 & Trust4 & .874 & 3.28 (1.16) &  &  &  \\
\midrule
\textbf{Subjective Norms (SN)} & SN1 & .772 & 3.42 (1.10) & \multirow{4}{*}{.906} & \multirow{4}{*}{.935} & \multirow{4}{*}{.783} \\
 & SN2 & .910 & 3.21 (1.13) &  &  &  \\
 & SN3 & .920 & 3.02 (1.19) &  &  &  \\
 & SN4 & .928 & 3.12 (1.15) &  &  &  \\
\midrule
\textbf{Perceived Enjoyment (PE)} & PE1 & .903 & 3.28 (1.14) & \multirow{4}{*}{.915} & \multirow{4}{*}{.940} & \multirow{4}{*}{.797} \\
 & PE2 & .855 & 3.18 (1.27) &  &  &  \\
 & PE3 & .924 & 3.25 (1.11) &  &  &  \\
 & PE4 & .888 & 3.58 (1.09) &  &  &  \\
\bottomrule
\end{tabular}}
\end{table}
\end{document}